\documentclass[12pt,preprint]{emulateapj}

\newcommand{\be} {\begin{equation}}

\def\uu {4U\,0142$+$614}

\def\kes {1E\,1841$-$045}

\def\rxj {1RXS\,J170849$-$400910}
\def\ea {1E\,2259$+$586}
\def\sgra{SGR\,1900+14}
\def\sgrb{SGR\,1806--20}

\newcommand{\bc}{\begin{center}}
\newcommand{\ec}{\end{center}}
\def\ltsima{$\; \buildrel < \over \sim \;$}
\def\lsim{\lower.5ex\hbox{\ltsima}}
\def\loe{\lower.5ex\hbox{\ltsima}}
\def\gtsima{$\; \buildrel > \over \sim \;$}
\def\gsim{\lower.5ex\hbox{\gtsima}}
\def\goe{\lower.5ex\hbox{\gtsima}}

\input psfig.sty

\shortauthors{REA ET AL.}
\shorttitle{MODELLING 4U0142+61 SPECTRUM}

\begin{document}
\title{Spectral modelling of the high energy emission of the magnetar 
4U\,0142+614}

\author{ N. Rea\altaffilmark{1,2},  R. Turolla\altaffilmark{3}, S. Zane\altaffilmark{4},  A. Tramacere\altaffilmark{5}, L. Stella\altaffilmark{6}, G.L. Israel\altaffilmark{6}, R. Campana\altaffilmark{5}}
\altaffiltext{1}{SRON Netherlands Institute for Space Research,
Sorbonnelaan, 2, 3584CA, Utrecht, The Netherlands; N.Rea@sron.nl}
\altaffiltext{2}{University of Sydney, School of Physics A29, NSW 2006, Australia}
\altaffiltext{3}{University of Padua, Department of Physics , via
Marzolo 8, 35131 Padova, Italy}
\altaffiltext{4}{MSSL, University
College London, Holmbury St. Mary, Dorking Surrey, RH5 6NT, UK}
\altaffiltext{5}{University of Rome, La Sapienza, Physics
Department, Piazzale A. Moro 2, 00185 Rome, Italy}
\altaffiltext{6}{INAF--Astronomical Observatory of Rome, via
Frascati 33, 00040, Monteporzio Catone (RM), Italy}

\begin{abstract} 

  We present an empirical spectral modelling of the high energy
  emission of the anomalous X-ray pulsar 4U 0142+614, based on
  simultaneous Swift and INTEGRAL observations from X to $\gamma$-ray
  energies. We adopted models contained in the XSPEC analysis package,
  as well as models based on recent theoretical studies, and
  restricted ourselves to those combinations of up to three components
  which produce a good fit while requiring the lowest number of free
  parameters. Only three models were found to fit satisfactorily the
  0.5--250~keV spectrum of 4U 0142+614: i) a $\sim 0.4$\,keV blackbody
  and two power-laws, ii) a resonant cyclotron scattering model plus a
  power-law and iii) two log-parabolic functions. We found that only
  the latter two models do not over-predict the infrared/optical
  emission observed simultaneously from this AXP, and only the
  log-parabolic functions can naturally account for the upper limits
  set by COMPTEL in the $\gamma$-ray range.  A possible interpretation
  of the two log-parabolae in terms of inverse Compton scattering of
  soft X-ray photons by very energetic particles is discussed.
\end{abstract}



\section{INTRODUCTION}

Soft $\gamma$-ray Repeaters (SGRs) and Anomalous X-ray Pulsars (AXPs)
are peculiar classes of high energy sources which share a number of
properties (among others, the emission of short bursts, slow rotation
periods in a narrow range, secular spin down and a X-ray luminosity of
$\approx10^{34}$--$10^{36}$\,erg~s$^{-1}$, see e.g. Woods \& Thompson
2006 for a recent review).  SGRs and AXPs are believed to be powered
by ultra-magnetic neutron stars (``magnetars"; Duncan \& Thompson
1992; Thompson \& Duncan 1995).

Spectral analysis is an important tool in magnetar astrophysics since
it can provide key information on the emission mechanism(s).  The first
attempts at modelling the spectra of AXPs were based on soft X-ray
data ($<10$~keV) only. It was found that 
a blackbody ($kT\sim$0.3--0.6\,keV) plus a power-law (photon index
$\Gamma\sim$2--4) or, limited to a few cases, two blackbodies ($kT_1
\sim$0.1 and $kT_2\sim$0.8\,keV) could successfully 
model the 0.5--10\,keV emission.  Even though the blackbody plus
power-law spectral model has been frequently applied to the X-ray
spectra of magnetar candidates, a convincing physical interpretation
is still missing (for the power-law component in particular).

Recent studies have attempted to overcome this limitation. Thompson,
Lyutikov \& Kulkarni (2002) proposed that magnetars have twisted
magnetospheres threaded by currents, which can efficiently boost via
resonant cyclotron scattering the soft thermal photons emitted by the
star surface.  Recent investigations (Lyutikov \& Gavriil 2006;
Fernandez \& Thompson 2006) showed that such a process may lead to the
formation of the power-law tail observed in the soft X-rays. Rea et
al.~(2006,~2007) presented an application of the Lyutikov \& Gavriil
(2006) resonant cyclotron scattering model (RCS model in the
following) to SGRs and AXPs, showing that in a few sources this
component alone can explain the X-ray spectrum below $\sim 10$\,keV.

More recently, hard X--ray observations have shown that in some
magnetar candidates (namely \uu, \rxj, \kes, \ea, \sgrb, \sgra;
Kuiper, Hermsen \& M\'endez~2004; Kuiper et al.~2006; Mereghetti et
al. 2005; G\"otz et al.~2006) a large fraction of the total flux is
emitted at energies well above $\sim10$~keV. The origin of these
``hard'' tails, that have so far been detected up to energies of $\sim
250$~keV, is still debated (see Beloborodov \& Thompson 2006 and Baring \&
Harding 2006 for possible interpretations).

AXPs and SGRs have been discovered to emit also at optical and/or NIR
wavelengths (e.g. Hulleman et al.~2000; Israel et al.~2004; Durant \&
van~Kerkwijk 2006b).  Their optical/NIR flux represents a small
fraction of the bolometric flux, but still can place important
constraints on models.

Several AXPs, have exhibited long term variability both in their
optical/infrared emission and in the X-rays (Oosterbroek et al. 1998;
Israel et al. 2002; Hulleman et al. 2004; Rea et al.~2005).  Unavoidably 
this introduces additional uncertainties in the modelling of broad band
spectra, based on observations at different wavelengths taken at
different times.

The AXP 4U\,0142+61 has been recently observed quasi-simultaneously in
the hard X-rays (with INTEGRAL, see den Hartog et al.~2006), soft
X-rays (with Swift--XRT, see \S\ref{obs}), radio, NIR and Optical
(with the Westerbork Synthesys Radio Telescope and Gemini North; see
den Hartog et al.~2006 and Durant \& van~Kerkwijk 2006b for all
details).

\begin{figure*}
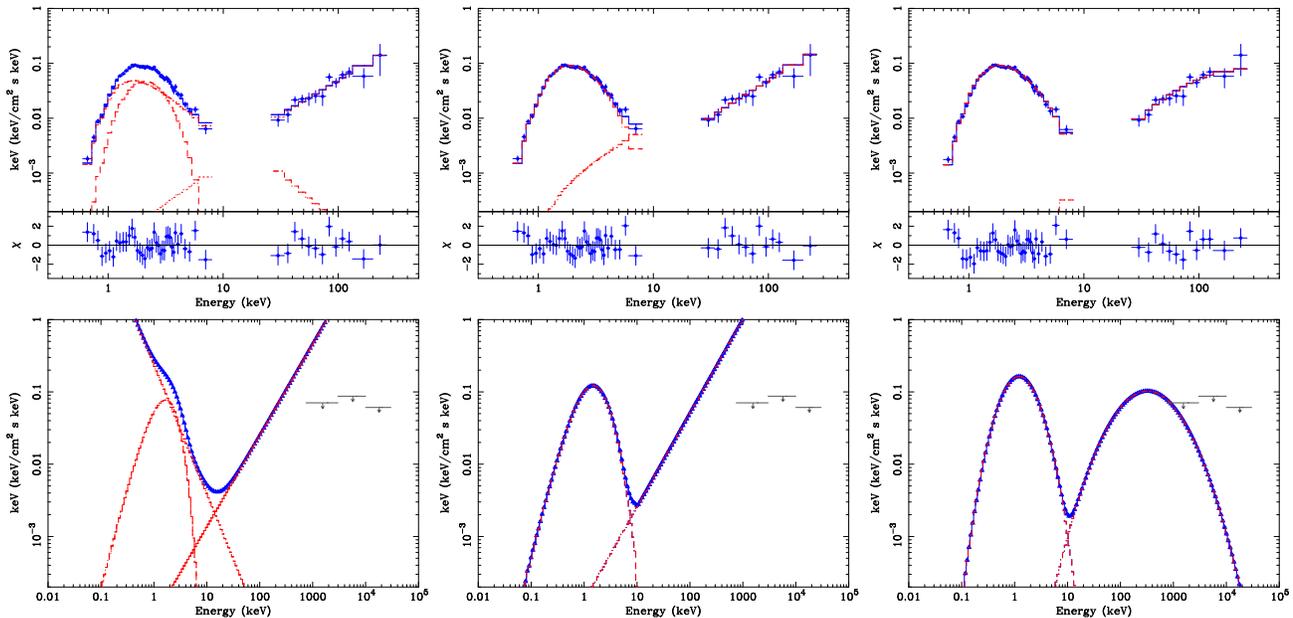

\hbox{
\vbox{
\psfig{figure=f1.ps,angle=270,width=5.5cm}
\psfig{figure=f2.ps,angle=270,width=5.5cm}
\psfig{figure=f3.ps,angle=270,width=5.62cm,height=4cm}}
\vbox{
\psfig{figure=f4.ps,angle=270,width=5.5cm}
\psfig{figure=f5.ps,angle=270,width=5.5cm}
\psfig{figure=f6.ps,angle=270,width=5.62cm,height=4cm}}
\vbox{
\psfig{figure=f7.ps,angle=270,width=5.5cm}
\psfig{figure=f8.ps,angle=270,width=5.5cm}
\psfig{figure=f9.ps,angle=270,width=5.62cm,height=4cm}}}
\caption{ \uu\, soft and hard X--ray spectrum fitted with (from left to
right): a blackbody plus two power--laws (\S~3.1), the RCS model plus
a power--law (\S~3.2), and two log--parabolic functions
(\S~3.3). Upper rows show the unfolded spectrum and the residuals of
the fits: in blue the data, the residuals and the overall model and in
red the single components. Lower row present the relative unabsorbed
fitted models: in red the single components, in blue the overall model
and the grey arrows report the COMPTEL upper limits from Kuiper et
al.~2006.}
\end{figure*}


Data from this and other campaigns show that most of the bolometric
emission ($>95$\%) revealed so far from this object lies in the
X/$\gamma$--ray energy range.  Motivated by this we present here an
empirical modelling of the 0.5--250\,keV spectrum of 4U\,0142+614
obtained during this quasi-simultaneous campaign (see \S\ref{spectra}).

In \S\ref{discussion} we discuss how the extrapolation of different
models match the flux measured at other wavelengths. In particular, we
note that only the fit with two absorbed log-parabolic functions is
consistent with the \uu\, $\gamma$-ray upper limits and does not
overestimate the observed optical/infrared emission. Some physical
implications of our study are then briefly discussed.


\begin{table*}
\begin{center}
\caption{Parameter values of the \uu\, best spectral models}
\begin{tabular}{lrlrlr}
\tableline\tableline
\multicolumn{2}{c}{BB + 2PL} & \multicolumn{2}{c}{RCS + PL} & \multicolumn{2}{c}{2 log-parabolae} \\
\tableline
$N_{H}$    & $0.96^{+0.08}_{-0.04}$ &  & $0.63^{+0.01}_{-0.04}$      &  & $ 0.73^{+0.1}_{-0.1}$   \\ 

constant &  0.96    & & 0.90 & &  0.87  \\ \\

kT\,(keV) & $0.421^{+0.01}_{-0.002}$ & kT\,(keV)  &   $0.335^{+0.002}_{-0.003}$   &  $E_{p1}$\,(keV) & $ 1.32^{+0.05}_{-0.05}$  \\

BB\,Flux  &  $ 1.53^{+0.07}_{-0.04}$  &  $\tau_0$ &  $1.84^{+0.19}_{-0.06}$ &    $\beta_1$ & $-2.52 ^{+0.08}_{-0.04}$  \\

$\Gamma_{\rm soft}$      &  $3.77^{+0.1}_{-0.01}$ & $\beta_{th}$ &  $0.22^{+0.02}_{-0.02}$  &  $logP_{1}$~Flux  & $ 3.9^{+1.1}_{-1.0}$ \\

$\Gamma_{\rm hard}$      & $0.73^{+0.08}_{-0.05}$ & $\Gamma_{\rm hard}$ &  $0.67^{+0.05}_{-0.03}$  &  $E_{p2}$\,(keV) & $412^{+43}_{-40}$ \\

PL$_{\rm soft }$\,Flux  &  $7.8^{+1.7}_{-1.7}$  & RCS\,Flux &  $3.2^{+0.3}_{-1.7}$  &   $\beta_2$ & $-0.7^{+0.1}_{-0.2}$  \\

PL$_{\rm hard }$\,Flux  &  $ 1.8^{+0.4}_{-1.2}$ & PL$_{\rm hard}$\,Flux &  $1.9^{+0.8}_{-1.1}$ &  $logP_{2}$~Flux & $ 1.8^{+2.2}_{-1.0}$ \\ \\

Total Abs. Flux     & $2.7^{+0.5}_{-0.5}$  &  &  $2.4^{+0.6}_{-1.0}$ &  & $3.5^{+0.4}_{-1.0}$     \\

Total Flux         & $9.7^{+0.6}_{-1.0}$    &  &  $10.3^{+0.5}_{-1.5}$ &  & $5.7^{+1.4}_{-1.0}$ \\

$\chi_{\nu}^2$ (d.o.f.)     & 1.04 (48) &  & 1.01 (48) & & 1.00 (48)      \\

\tableline\tableline
\end{tabular}
\tablecomments{Best fit parameters for the 0.5--250\,keV spectral modelling of \uu. Errors are at
1$\sigma$ confidence level. Fluxes (if not otherwise specified) are
unabsorbed and in units of $10^{-10}$\,erg\,cm$^{-2}$\,s$^{-1}$. RCS,
BB, PL$_{\rm soft}$ and $logP_{1}$ fluxes are for the 0.5--10\,keV
energy band, while PL$_{\rm hard}$ and $logP_{2}$ fluxes refer to the
20--250\,keV band.  Total fluxes are in the 0.5--250\,keV
band. $N_{H}$ is in units of $10^{22}$\,cm$^{-2}$. The {\it constant}
parameter, which accounts for the intercalibration, assumes Swift--XRT
as a reference.}

\end{center}
\end{table*}


\section{The observations}
\label{obs}

We used Swift X-ray Telescope (XRT) data obtained on 2005 July
11--12th for an on--source exposure time of 7400\,s in Photon Counting
(PC) mode (that produced a standard 2D image). The data were analysed
with the FTOOL task {\tt xrtpipeline} (version build-14 under HEADAS
6.0). We applied standard screening criteria to the data, and hot and
flickering pixels, high background intervals and bright limb were
removed. Screened event files were then used to derive light curves
and spectra.  We included data between 0.5--8\,keV, where the PC
response matrix is well calibrated (we used the latest v.8 response
matrices). We extracted photons from an annular region (3 pixels inner
radius, 30 pixel outer radius) in order to avoid pile-up contamination
and we considered standard grades for the PC mode 0--12.

For the hard X-ray part of the spectrum, we used INTEGRAL data taken
by the IBIS/ISGRI instrument (20--250\,keV energy range)
quasi-simultaneously with the Swift XRT ones. These consist of 868\,ks
divided in 265 Science Windows, collected between 2005 June 29th and
July 17th, in a pointed observation (see den Hartog et al.~2006).

\section{SPECTRAL MODELLING}
\label{spectra}

We fitted (making use of XSPEC 11.3 and 12.1) the X/$\gamma$--ray
spectrum of \uu \ by using any combination of the following
components: a) for the soft X-ray part -- two blackbodies, a blackbody
plus a power-law, a neutron star atmosphere model ({\tt nsa} model,
only available for a maximum magnetic field of B=$10^{13}$\,G), a
comptonized blackbody ({\tt compbb}), a disk blackbody ({\tt diskbb}),
a log-parabolic function, a RCS model (as in Rea et al.~2006,~2007),
and b) for the hard X-ray part -- a power-law, a bremsstrahlung model
({\tt bremss}), a log-parabolic function. In all cases we included
photoelectric absorption (abundances set to solar values from Lodders
2003) and a constant to account for cross-calibration uncertainties in
the two instruments.

All combinations, except for the three models discussed below (see
also Fig.\,1, 2, and Tab.\,1), gave unsatisfactory reduced $\chi^2$
values ($\chi_{\nu}^{2} > 1.8$). All the three models that successfully
reproduce the spectrum require 8 free parameters. Of course, we
found that other models, involving components not listed above (e.g. a
cut-off power-law instead of the hard X-ray power-law), gave
satisfactory results. However, in all cases more than 8 free parameters
were required.

\subsection{The canonical empirical model}

This is the empirical model which is usually adopted for AXP spectra
and consists of a single blackbody and two power-laws. For \uu\, the
best fit gave a blackbody temperature of $kT\simeq0.42$\,keV, a very
steep power-law below $\sim10$~keV with $\Gamma\simeq 3.77$, and a
flatter power law for the high-energy part of the spectrum,
$\Gamma\simeq0.73$ (see also Kuiper et al.~2006; den~Hartog et
al.~2006). We tried also a slightly more general model which includes
a cut-off around 300\,keV in the attempt to make the extrapolation of
the canonical model compatible with the existing upper limits above
$\sim 1$~MeV (note that we did not include the MeV upper limits in the
fit).  We found that the quality of the fit did not improve much,
despite the addition of one more free parameter in this
modelling. Note that the model consisting of two blackbodies plus a
power-law did not provide a good fit, mainly because the two
blackbodies alone cannot account for the soft X-ray part of the
emission.

\subsection{The Resonant Cyclotron Scattering model}

This combination makes use of the RCS model, based on the spectra
computed by Lyutikov \& Gavriil (2006) and recently implemented in
XSPEC by Rea et al.~(2006). As shown by Rea et al.~(2006,~2007), this
component can explain well the soft X-ray emission of a few magnetars,
whereas an additional power-law is required for sources of this class
that have a strong hard X-ray emission. The model parameters, together with the
ranges over which each parameter was allowed to vary, are: the
velocity of the magnetospheric $e^-$ currents (in units of $c$)
$0.1<\beta_{th}<0.5$; their (Thomson) scattering optical depth, $1 <
\tau_0 < 10$; and the temperature of the seed 
surface emission (assumed to be a blackbody), $0.1$~keV$<kT<3$~keV.

The efficiency of RCS drops above $\sim 10$\,keV, but even in the
6--10\,keV energy range the addition of another model component is
required to fit the X-ray spectrum of \uu. We find a best fit of the
whole 0.5--250~keV spectrum with a RCS plus power-law model with a
surface temperature of $kT\simeq0.33$\,keV, magnetospheric plasma
parameters of $\beta_{th}\simeq0.22$ and $\tau_0 \simeq1.84$, and a
power law index of $\Gamma\simeq0.67$. The RCS component dominates the
emission below 6~keV.

\subsection{The log-parabolic model}

The log-parabolic model is an empirical model commonly used to fit
blazars spectra, but also successfully applied to a few radio pulsars
displaying hard X-ray emission, e.g. the Crab and the Vela pulsars
(Kuiper et al.~2001; Massaro et al.~2006a,b). We applied this model to
our data and found that it fits well the soft and hard X-ray spectrum
of \uu\, (see also Kuiper et al. 2006).  Log-parabolic spectra (Landau
et al.~1986) can be obtained when relativistic electrons are
accelerated by a statistical acceleration mechanism in which the
probability of acceleration depends on energy, and cool via
synchrotron losses (Landau et al.~1986; Massaro et al.~2004) or by
inverse Compton scattering of synchrotron seed photons.

The log-parabolic spectral distribution can be regarded as the
simplest generalization of a power-law distribution, as it comprises a
mild symmetric spectral curvature. We adopted the following
expression, $F(E) = KE^{-\alpha -\beta \log E}$, where $E$ is the
energy in keV, $\alpha$ is the photon index at 1~keV energy and
$\beta$ controls the curvature of the parabola in a log-log
representation. The log-parabola peaks at $E_p =
10^{(2-\alpha)/2\beta}$; its free parameters, besides the
normalization constant $K$, are $\alpha$ and $\beta$ or, equivalently,
$E_p$ and $\beta$.

The two log-parabolae we fitted have peaks at $\simeq$ 1.32 and
412\,keV, and curvatures of $\beta=$--2.52 and --0.7.  They provide a
very good fit to the data, and have the same number of free parameters
as the models discussed above (see Fig.~1 and Tab.\,1).

\section{DISCUSSION and INTERPRETATION}
\label{discussion}

We have shown that among spectral models with a relatively limited
number of free parameters ($N_{par}=8$), only three successfully fit
the X-ray spectrum of \uu\ (see \S\ref{spectra}).

The 0.5--250~keV fluxes we derived, are consistent among all the three
best models. $N_{H}$ in the RCS and log-parabolic models is closer to
the value measured for this source using X-ray edges in grating
spectra ($\sim6\times10^{21}$\,cm$^{-2}$; see Durant \& van~Kerkwijk
2006a).

Comparing the extrapolations of these models with the radio, infrared
and optical fluxes measured from quasi-simultaneous observations
(Durant \& van~Kerkwijk~2006b, den~Hartog et al.~2006), as well as
$\gamma$-ray upper limits obtained around 0.2~MeV by COMPTEL (Kuiper
et al.~2006), we found that the canonical model (\S\,3.1) overpredicts
all of them.  The RCS model (\S\,3.2) and the two log-parabolae
(\S\,3.3) improve over the canonical model since they do not
overpredict the low energy emission. In any case, the optical/NIR
emission of 4U0142+614 (and likely of other magnetar candidates)
accounts for only less than 5\% of the overall source luminosity, and
might be due to secondary processes as compared to those responsible
for the X/$\gamma$--ray flux. The optical emission is pulsed, so it is
likely to be non-thermal in origin (Kern \& Martin 2000). On the
contrary, the IR emission can be ascribed to a fossil disk (see
e.g. Wang et al. 2006), reprocessing a fraction of the soft/hard X-ray
emission of the magnetar.

The two log-parabolae (\S\,3.3) are the only model that naturally
accounts for the fast decline of the spectrum in the $\gamma$-rays,
implied by the COMPTEL upper limits.  Nevertheless, in the other two
models these limits would not be violated if the hard X-ray power law
were characterised by a cut-off at $\sim300$\,keV (though this would
require an additional free parameter in the fit).  Note that in a more
detailed physical model the high-energy power law required to fit the
INTEGRAL data may quickly decline around $\sim300$~keV, e.g. as a
result of the maximum energy up to which the spectrum of the emitting
particles extends.


\begin{figure}[t]
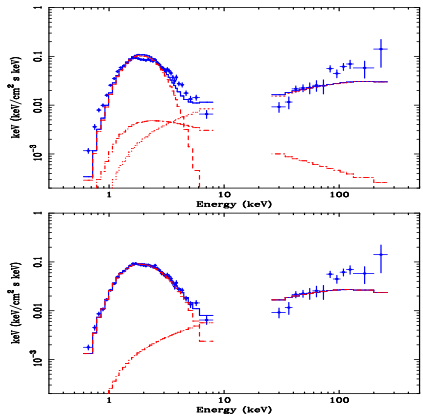

\centerline{
\vbox{
\psfig{figure=f2a.ps,angle=270,height=2.7cm,width=5.5cm}
\psfig{figure=f2b.ps,angle=270,height=2.7cm,width=5.5cm} } }
\caption{\uu\, soft and hard X--ray spectrum fitted with an absorbed blackbody plus a power-law and a bremsstrahlung (top panel) and an absorbed resonant scattering model plus a bremsstrahlung (bottom panel). Both models give an unsatisfactory reduced $\chi^2> 2$.}

\end{figure}


Thompson \& Belobodorov (2005) recently proposed that the hard X-ray
magnetar emission may be attributed to bremsstrahlung photons emitted
by a thin surface layer heated up to $\sim 100$~keV by returning
currents. However, we find that the \uu \ spectrum is harder at high
energies than envisaged this scenario (Fig.\,2). A bremsstrahlung
model seems hardly compatible with the hard X-ray spectrum
of this source and the COMPTEL upper limits.

Although two log--parabolic functions are often used to model a
Synchrotron Self Compton (SSC) emission, in the case at hand this
interpretation appears unlikely for several reasons. First, it is
difficult to reconcile the soft part of the X-ray spectrum with
synchrotron emission, since the low energy log-parabola has a
curvature $\beta=-2.52$ which translates in a curvature of $\sim 12$
for the e$^-$ spectrum. This is very close to the cut-off curvature of
the high energy synchrotron emission of a single particle. Moreover,
we performed detailed simulations (see Tramacere 2002; Tramacere \&
Tosti 2003 for details) in order to infer which local magnetic field
and electron density is needed to power the \uu\, emission by a SSC
mechanism. We obtained $B\sim10^8$\,G and $n_e\sim
10^{15}$\,cm$^{-3}$, respectively. A charge density of the same order
is predicted in the simple twisted magnetosphere model by Thompson,
Lyutikov \& Kulkarni (2002), but close to the neutron star
surface. Instead, the low value of the magnetic field requires that
such e$^-$ are located at an altitude of several hundred kms in the
magnetosphere.  We may speculate that inverse Compton scattering by a
population of relativistic particles high up in the magnetosphere
may explain the hard X--ray emission of magnetars, for other (than
synchrotron) spectral distributions of seed photons, e.g. that
produced by the RCS model. Interestingly, the presence of highly
relativistic particles well above the star surface has been also
suggested by Thompson \& Beloborodov (2005). These authors pointed out
that seed positrons injected at $\sim 100$~km above the surface can
undergo runaway acceleration and upscatter X-ray photons above the
threshold for pair production. In their model, the $e^\pm$ pairs would
then be responsible for efficient synchrotron emission in the 20-200
keV band.

In this Letter we introduced and fitted spectral models, as purely
empirical laws. As in many other studies where the total emission of
X--ray pulsars is modelled, no attempt was made to fit separately the
spectral behaviour of the pulsed and the non--pulsed components. A
more extensive study based on data of other magnetar candidates and
attempting a more detailed physical interpretation of these results is
in progress and will be reported elsewhere.

\vspace{.05cm}


NR acknowledges support from an NWO-Postdoctoral Fellowship and a
Short Term Visiting Fellowship awarded by The University of Sydney,
and SZ thanks PPARC for financial support. This work was partially
supported through ASI and MIUR funds. NR thanks S. Campana and
V. Mangano for a preliminary help with the Swift analysis,
M. van~Kerkwijk, M. Durant, W. Hermsen, L. Kuiper, P. den~Hartog,
M. M\'endez, J. Schuurmans, M. Baring, L. Burderi, S. Markoff and
E. Massaro for useful discussions. We also thank N. Gehrels for
approving the Swift observation within the Director Time and the
referee, Feryal Ozel, for useful suggestions.

\end{document}